\documentclass[letterpaper, 10 pt, conference]{ieeeconf} 
\IEEEoverridecommandlockouts

\usepackage{enumitem}
\setlist{topsep=0pt, leftmargin=*}
\usepackage{algorithm}
\usepackage{algorithmicx}

\usepackage[noend]{algpseudocode}
\usepackage{amsmath, amsfonts, bm, amssymb, tabularx}
\usepackage{latexsym, mathtools}
\usepackage{amsthm}
\usepackage{ifthen}
\usepackage{hyperref}\hypersetup{colorlinks=true, unicode=true, linkcolor=[rgb]{0.10,0.05,0.67}, citecolor=[rgb]{0.10,0.05,0.67}, filecolor=[rgb]{0.10,0.05,0.67}, urlcolor=[rgb]{0.10,0.05,0.67}}
\usepackage{Shorthands}
\usepackage{cleveref}
\newcommand{\citep}[1]{\cite{#1}}
\usepackage{stackengine}

\usepackage[numbers, sort&compress]{natbib}
\allowdisplaybreaks

\title{\LARGE \bf
Learning with Imperfect Models: \\ When Multi-step Prediction Mitigates Compounding Error
}
\usepackage{subcaption}
\author{
Anne Somalwar$^{1}$, Bruce D. Lee, George J. Pappas, Nikolai Matni
\thanks{$^1$ All authors are with the University of Pennsylvania. Emails : \tt\small\{somalwar, brucele, nmatni, pappasg\}@seas.upenn.edu}%
}

\begin{document}

\twocolumn

\maketitle
\thispagestyle{empty}
\pagestyle{empty}

\begin{abstract}
Compounding error, where small prediction mistakes accumulate over time, presents a major challenge in learning-based control. For example, this issue often limits the performance of model-based reinforcement learning and imitation learning. One common approach to mitigate compounding error is to train multi-step predictors directly, rather than relying on autoregressive rollout of a single-step model. However, it is not well understood when the benefits of multi-step prediction outweigh the added complexity of learning a more complicated model. In this work, we provide a rigorous analysis of this trade-off in the context of linear dynamical systems. We show that when the model class is well-specified and accurately captures the system dynamics, single-step models achieve lower asymptotic prediction error. On the other hand, when the model class is misspecified due to partial observability, direct multi-step predictors can significantly reduce bias and thus outperform single-step approaches. These theoretical results are supported by numerical experiments, wherein we also (a) empirically evaluate an intermediate strategy which trains a single-step model using a multi-step loss and (b) evaluate performance of single step and multi-step predictors in a closed loop control setting. 
\end{abstract}
\section{Introduction}

A typical approach to time series forecasting is to fit a one-step ahead prediction model and apply it recursively to obtain predictions over multiple time steps. In doing so, small errors may compound over time, leading to poor long-horizon prediction. This issue hinders the application of such single-step models in e.g., controller design.

By directly training multi-step models to predict longer horizons, the issue of compounding error can be mitigated. The main drawback of doing so is that the number of parameters for a direct multi-step predictor scales with the prediction horizon, thus potentially requiring more data to achieve a desired prediction performance. While this tradeoff between prediction horizon accuracy and data requirements is broadly known to exist, 
 it is primarily studied from an empirical perspective \citep{lambert2021learning, lambert2022investigating}. We therefore lack principled guidance for exactly when direct multi-step prediction should be preferred over autoregressive rollout of single-step models. Motivated by this challenge, we provide a rigorous comparison of the sample efficiency of learning multi-step predictors with that of learning single-step predictors in the 
 setting of a linear dynamical system.

\subsection{Related Work}

\paragraph{Multi-step Identification}
The goal of system identification is to use data to learn a model that can be used for forecasting or control \citep{ljung1998system}. 
To this end, one typically wants to select a model from the hypothesis class that minimizes the simulation error, i.e., the cumulative prediction error over all future time steps. Due to the computational challenge of doing so, it is much more common to instead learn a model which minimizes the single-step prediction error and apply it autoregressively \citep{farina2008some}. However, such approaches tend to generalize poorly if the underlying data generating process does not belong to the hypothesis class. 
This has motivated the application of algorithms such as Data as Demonstrator (DaD) which use approaches from imitation learning to train a predictor to self-correct its prediction error at each step, leading to improved empirical performance \citep{venkatraman2015improving, venkatraman2017improved}. 

Direct learning of prediction models for each time-step in the prediction horizon partially bypasses the issue of compounding error, and has proven successful for both model predictive control \citep{terzi2018learning, balim2024stochastic, kohler2022state} and value approximation in MDPs \citep{asadi2019combating}. Empirical studies of single-step and multi-step dynamics models learned with various neural network architectures have also been conducted. Namely, \citet{lambert2022investigating} investigate the performance of recursive application of single-step models parameterized by neural networks in a handful of examples and characterize circumstances which may worsen the effects of compounding error. \citet{chandra2021evaluation} give a comparison of various deep learning architectures for predicting multiple steps of a time series and show the efficacy of  bidirectional and encoder-decoder LSTM networks.

These empirical studies underscore the importance of careful consideration when designing a model for multi-step prediction. Our work studies this question in stylized settings, allowing us to clearly demonstrate the potential drawbacks and benefits of direct multi-step prediction as compared to more traditional single-step approaches. 

\paragraph{Learning-Enabled Control} 

While the issue of compounding error has long been studied in system identification and control, it has resurfaced as a prominent issue in the learning community, exacerbated by the use of neural networks as function approximators. In model-based reinforcement learning, it has been observed that synthesized controllers may exploit compounding errors of the learned model \citep{chua2018deep,levine2020offline}, motivating numerous heuristics for accounting for the model error during policy synthesis \citep{janner2019trust, yu2020mopo, yu2021combo, kidambi2020morel}. \citet{lambert2021learning} instead propose learning a direct multi-step predictor parametrized by a low dimensional decision variable which may be optimized online. The issue of a mismatch between the hypothesis class and the underlying data generating process also poses a challenge for behavior cloning. For example, incorrectly assuming that the demonstrator is Markovian can lead to a policy that deviates substantially from the demonstrator \citep{chi2023diffusion, block2023provable}. This can be remedied in part by replacing the standard behavior cloning objective with an objective that predicts the expert actions for multiple timesteps to maintain temporal consistency, resulting in so-called \emph{action-chunking} based approaches~\citep{block2023provable, zhao2023learning}. We draw inspiration from these studies, and consider instances of linear systems with either Markovian or non-Markovian observations to compare direct multi-step prediction and autoregressive evaluation of single-step predictors.

\subsection{Contributions}

We provide the first quantitative comparison of the multi-step prediction error incurred by directly learned multi-step predictors against that incurred by autoregressive evaluation of a single-step predictor. In particular:
\begin{itemize}[noitemsep,nolistsep]
    \item We provide an asymptotic characterization of the multi-step prediction error for the two methods in the setting of a fully observed dynamical system. Our results show that for stable systems with a small spectral radius, the prediction error of autoregressive evaluation decays significantly faster with increasing data than that of direct multi-step prediction. This benefit diminishes as the spectral radius increases to one.
    \item We characterize the prediction error for the two methods applied to a partially observed dynamical system which is incorrectly assumed to be fully observed by the hypothesis class, thus addressing the issue of trajectory prediction under misspecification due to an unjustified Markovian assumption. These results demonstrate that multi-step predictors may enjoy significantly lower bias in the face of model misspecification.
    \item We conduct numerical experiments that compare the aforementioned approaches with an additional standard baseline: fitting a single-step model that minimizes the multi-step prediction error. 
    \item We empirically evaluate the performance of autoregressive rollouts of single step predictors and multi-step predictors in a closed loop control setting. 
\end{itemize}
Our results, while limited to stylized settings, capture key properties of real systems---such as partial observability and misspecification---and thus provide useful guidance to practitioners navigating the complex design space of data-driven multi-step predictors.\\ 

\noindent
\textbf{Notation}: $\mathcal{N}(\mu, \sigma^2)$ denotes a normal distribution with mean $\mu$ and variance $\sigma^2$, $\rho(\cdot)$ denotes the spectral radius of a matrix, $\norm{\cdot}$ denotes the vector Euclidean norm, $\norm{\cdot}_F$ denotes the matrix Frobenius norm, $\VEC(\cdot)$ denotes the vectorization of a matrix, and $\otimes$ denotes the Kronecker product. 


\section{Problem Formulation}

Consider the linear time invariant dynamical system 
\begin{equation}
\label{eq: dynamics}
\begin{aligned}
    x_{t+1} &= Ax_t + Bu_t + B_w w_t &&\quad t \in \mathbb{Z}^+\\
    y_t &= C x_t + D_v v_t,  &&\quad t \in \mathbb{Z}^+
\end{aligned}
\end{equation}
with state $x_t \in \R^{\dx}$, input $u_t \in \R^{\du}$, observation $y_t\in R^{\dy}$, process noise $w_t \overset{\iid}{\sim} \calN(0,I_{d_x})$, sensor noise $v_t  \overset{\iid}{\sim} \calN(0,I_{d_y})$, and initial condition $x_0 = 0$. We assume that $(A,C)$ is observable, that $(A, \bmat{B, B_w})$ is controllable, $\rho(A) < 1$, and that the control inputs are selected randomly as $u_t\overset{\iid}{\sim}\calN(0,I_{d_u})$.  

We assume that the dynamics \eqref{eq: dynamics} are unknown, and our goal is to learn a predictor that forecasts a horizon $H$ of future observations using past observations. To this end, we suppose that we are given a dataset $\calD_N = \curly{(y_t, u_t)}_{t=1}^N$ 
collected from a training rollout of \eqref{eq: dynamics} which will be used to determine a function $\hat f_H$ belonging to a hypothesis class $\calF_H$. This function will be used to predict $y_{t+1:t+H}$ given $y_{1:t}$ and $u_{1:t+H-1}$. 

The quality of the learned function will be measured by the loss
\begin{align}
    \label{eq: loss}
    L(\hat f_H) \triangleq  \bar \E \norm{y_{t+1:t+H} - \hat f_H(y_{1:t}, u_{1:t+H-1})}^2,
\end{align}
where the operator $\bar \E$ is defined as
\begin{align*}
    \bar \E[f(t)] \triangleq \lim_{T\to\infty} \mathbf{E} \frac{1}{T}\sum_{t=0}^T f(t), 
\end{align*}
and the expectation is taken over an evaluation rollout of system \eqref{eq: dynamics} that is independent of the dataset $\calD_N$. This is equivalent to taking an expectation under the steady state distribution for the system \eqref{eq: dynamics}. 

To provide rigorous understanding of situations where multi-step prediction does or does not help, we consider a simplified setting in which the hypothesis class consists of static linear predictors, i.e., a function $f_H \in \calF_H$, is given by \begin{equation}\label{eq:gen-pred}
f_H(y_{1:t}, u_{1:t+H-1}) = G \bmat{y_t \\ u_{t:t+H-1}}, 
\end{equation}
for a matrix $G \in S \subseteq \R^{H \dy \times (\dy + H\du)}$.  Here the subspace $S$ encodes whether we are fitting a multi-step or single-step model: we provide explicit parameterizations for these model-classes in the next subsections. Our restriction to static linear predictors of the form~\eqref{eq:gen-pred} assumes that the observation sequence is Markovian, i.e., that a history of observations is unnecessary to predict the future trajectory.  In the sequel, we slightly abuse notation and denote the loss~\eqref{eq: loss} incurred by a predictor~\eqref{eq:gen-pred} defined by matrix $\hat G$ by $L(\hat G).$ 

We consider two settings: one where the Markovian assumption is justified ($C=I$ and $D_v = 0$), resulting in a well-specified problem, and one where it is not justified ($C \neq I$ and $D_v \succ 0$), resulting in a misspecified problem. 
In these two settings, we compare the $H$ step prediction error~\eqref{eq: loss} incurred by a learned single-step model rolled out for $H$ timesteps to that incurred by a directly learned $H$-step model. 

\subsection{Single-step Predictors}
\label{s:singlestep}
The single-step approach first solves
\begin{align}
    \label{eq: single-step LS}
    \bmat{\hat G_y & \hat G_u} = \argmin_{\substack{G_y \in \R^{\dy \times \dy} \\ G_u \in \R^{\dy \times \du}} } \sum_{t=1}^{N-1} \norm{y_{t+1} - \bmat{G_y & G_u} \bmat{y_t \\ u_t}}^2.
\end{align}

Using this model, one can predict $y_{t+1:t+H}$ by rolling out $\bmat{\hat G_y & \hat G_u}$ autoregressively:
\begin{align}
    \label{eq: single-step rollout}
    \hat y_{t+1} =&  \bmat{\hat G_y & \hat G_u} \bmat{y_t \notag \\ u_t} \\
    \hat y_{t+2} =& \bmat{\hat G_y & \hat G_u} \bmat{\hat y_{t+1} \\ u_{t+1}} \notag \\ 
    &\vdots \notag \\ 
    \hat y_{t+H} =& \bmat{\hat G_y & \hat G_u} \bmat{\hat y_{t+H-1} \\ u_{t+H-1}}. 
\end{align}
The resulting $H$-step predictor can be composed to form a direct mapping from the data to the predicted trajectory as 
\begin{align}
    \label{eq: single-step rolled out G}
    \hat G^{SS}_N = \bmat{\hat G_y & \hat G_u & 0 & \dots & 0 \\
                   \hat G_y^2 & \hat G_y \hat G_u & \hat G_u   &\dots & 0 \\
                   \vdots  &&& \ddots \\
                   \hat G_y^{H} &   \hat G_y^{H-1}\hat G_u &  \hat G_y^{H-2}\hat G_u &\dots & \hat G_u \\}.
\end{align}
As past predictions become part of the regressor for future predictions, this approach often suffers from compounding error. 

\subsection{Multi-step Predictors}
The issue of compounding error from autoregressive roll-out of a single-step model motivates direct multi-step approaches which directly minimize the $H$ step prediction error:
\begin{align}
    \label{eq: multistep LS}
    \hat G^{MS}_N = \argmin_{G \in S} \sum_{t=1}^{N-H} \norm{y_{t+1:t+H} - G \bmat{y_t \\ u_{t:t+H-1}}}^2,
\end{align}
for $S \subseteq \R^{H\dy \times (\dy + H\du)}$. 
We consider the function class which fits $H$ distinct predictors, one for each step in the prediction horizon. This amounts to setting $S = \R^{H \dy \times (\dy + H \du)}$.\footnote{One could impose the causality structure, i.e. that $S$ has a triangular structure. We refrain from doing so for simplicity, and due to the fact that future inputs are independent of the past.}

There is a tradeoff induced by fitting multi-step predictors rather than single-step predictors. In particular, the single-step predictor is subject to compounding error, while the complexity of the above identification problem increases for longer horizons. We study this tradeoff in the two aforementioned settings: a system with Markovian observations, and a system with non-Markovian observations. Due to the Markovian assumption for the identification problem, these cases serve as instances where the identification problem is well-specified and misspecified, respectively.

\subsection{Intermediate Formulations}

\stackMath
Rather than fitting independent predictors for every timestep, one can instead formulate a hypothesis class for multi-step prediction with lower complexity. In particular, one could impose additional structure on $S$, e.g. 
\begin{align}\label{eq: multistep loss} S = \curly{\bmat{ G_y &  \!\!\!\!\!G_u & 0 & \!\!\!\!\!\dots & \!\!\!\!\!0 \\
                    G_y^2 & \!\!\!\!\!G_y G_u & \!\!\!\!\!G_u   &\!\!\!\!\!\dots & \!\!\!\!\!0 \\
                   \vdots  &&& \ddots \\
                   G_y^{H} & \!\!\!\!\!  G_y^{H-1} G_u & \!\!\!\!\! G_y^{H-2} G_u &\!\!\!\!\!\dots & \!\!\!\!\!G_u \\} \Bigg \vert \raisebox{-8pt}{\stackon[1pt]{G_u \in \R^{\dy \times \du}}{G_y \in \R^{\dy \times \dy}}}}. \end{align} 
This consists of functions which take the form of a single-step predictor that is applied auto-regressively. In contrast to the single-step approach of \Cref{s:singlestep}, solving \eqref{eq: multistep LS} with this choice of $S$ consists of a multi-step loss function for a class of single-step predictors, a common approach to mitigate the compounding error issue \citep{farina2008some} without increasing the number of parameters that must be learned.
We study the loss \eqref{eq: loss} of the predictors \eqref{eq: multistep LS} fit with classes \eqref{eq: multistep loss} in numerical experiments and leave analytically characterizing the asymptotic prediction error for this predictor to future work. 
\section{Well-Specified Setting}

In this section, we study the well-specified setting in which the Markovian assumption is valid. In particular, we restrict system \eqref{eq: dynamics} to be a fully observed system by assuming that $C = I$ and $D_v = 0$ so $y_t = x_t$ for all $t$.  

To compare the single-step and multi-step approaches in this setting, we first observe that either predictor $\hat f_H$ is defined in terms of a linear map $\hat G$ applied to the vector $\bmat{x_t^\top & u_{t:t+H-1}^\top}^\top$. Therefore the loss \eqref{eq: loss} may be written
\begin{align}
    \label{eq: loss expanded}
    L(\hat G) = \bar \E \norm{x_{t+1:t+H} - \hat G \bmat{x_{t} \\ u_{t:t+H-1}}}^2. 
\end{align}
Rolling out the dynamics, we find that 
\begin{align*}
    x_{t+1:t+H} = G^\star  \bmat{x_t \\ u_{t:t+H-1}} + \Gamma_w w_{t:t+H-1},
\end{align*}
where 
\begin{align*}
    G^\star &=  \bmat{A & B \\ A^2&  AB & B \\ &\vdots && \ddots \\ A^H & A^{H-1}B && \dots & B},
\end{align*}
and
\begin{align*}
    \Gamma_w &= \bmat{B_w \\  AB_w & B_w \\ \vdots && \ddots \\ A^{H-1} B_w & \dots && B_w}.
\end{align*} 

Then expanding $x_{t+1:t+H}$ in equation \eqref{eq: loss expanded}, and using the independence of $w_{t:t+H-1}$ from $x_t$ and $u_{t:t+H-1}$, we conclude that
\begin{align*}
    L(\hat G) &= \bar\E \norm{(\hat G - G^\star) \bmat{x_t \\ u_{t:t+H-1}}}^2+\bar \E \norm{\Gamma_w w_{t:t+H-1}}^2 \\
    &= \norm{(\hat G - G^\star) \Sigma_z^{1/2}}_F^2 + \norm{\Gamma_w}_F^2,
\end{align*}
where $\Sigma_z = \bar \E \bmat{x_t \\ u_{t:t+H-1}}\bmat{x_t \\ u_{t:t+H-1}}^\top$ is the stationary covariance for the regressor. 
Consequently, the discrepancy between the single-step and multi-step predictors is contained in the term $\norm{(\hat G - G^\star) \Sigma_z^{1/2}}_F^2$. We study the behavior of this term asymptotically, where $\hat G$, or equivalently $\hat G_N$, is an operator learned on the dataset of size $N$.\footnote{We sometimes omit the subscript $N$ on $\hat G_N$ to ease notational burden.} In particular, we examine
\begin{align*}
    \lim_{N\to\infty} N \E\brac{\norm{(\hat G_N - G^\star) \Sigma_z^{1/2}}_F^2},
\end{align*} 
for the single-step and multi-step predictors $\hat G^{SS}_N$ and $G^{MS}_N$, respectively, where the expectation is taken over the dataset used to fit $\hat G_N$. 

The reducible error of the multi-step predictor is characterized by the following proposition. 

\begin{proposition}
    \label{prop: well specified multistep}
    The reducible asymptotic error of the multi-step predictor $\hat G^{MS}_N$ is given by 
    \begin{align*}
        &\lim_{N\to\infty} N \mathbf{E} \brac{\norm{(\hat G^{MS}_N - G^\star) \Sigma_z^{1/2}}_F^2}\\
        &=\trace\paren{\Gamma_w ((M_{MS} + H\du I_H) \otimes I_{\dx}) \Gamma_w^\top},
    \end{align*}
    where $M_{MS} \in \R^{H \times H }$ is the matrix with entry $(i,j)$ given by
    $
        M_{MS}^{ij} = \trace(A^{\abs{i-j}})$.
\end{proposition}

 The above result shows that the error decays asymptotically at a rate of $1/N$. The scaling is characterized by the trace expression, which represents the asymptotic covariance of the estimation error; importantly, it grows with the horizon $H$ (note the $(M_{MS} + H d_u I_H)$ term). The error of the single-step predictor is characterized below. 


\begin{proposition}
    \label{prop: single-step well specified}
    The asymptotic error of the single-step predictor $\hat G^{SS}_N$ is given by 
    \begin{align*}
        &\lim_{N\to\infty} N \mathbf{E} \brac{\norm{(\hat G^{SS}_N - G^\star) \Sigma_z^{1/2}}_F^2} \\
        &=\trace(\Gamma_w ((M_{SS} + \du I_H) \otimes I_{\dx}) \Gamma_w^T),
    \end{align*}
    where $M_{SS} \in \R^{H \times H}$ is the matrix with entry $(i,j)$ given by
    \begin{align*}
          \trace\paren{\paren{ I - \Sigma_x^{-1} \sum_{\ell=0}^{\min\curly{i,j}-2} A^{\ell}B_wB_w^\top (A^\ell)^\top} (A^{\abs{j-i}})^\top}.
    \end{align*}
\end{proposition}
Again, the error decays at a rate $1/N$. In contrast to the multi-step predictor, the asymptotic scaling of the single-step prediction error has the quantity $M_{SS} + \du I_H$ inside the trace. This means that the multi-step predictor suffers an extra factor of $H$ in the input term. Additionally the matrix $M_{MS}$ for the multi-step case has entries which decay as the distance to the diagonal increases, while $M_{SS}$ has entries which decay as the distance to the upper left element increases. Roughly, this  indicates that for very stable systems $M_{SS}$ should become smaller than $M_{MS}$. We make this concrete in the sequel.  


\subsection{Comparison}


We can express the quadratic form defining the reducible portion of the error in \Cref{prop: well specified multistep} as the reducible error in \Cref{prop: single-step well specified} plus the additional term $\trace (\Gamma_w((M_{MS} - M_{SS} + (H-1)d_uI_H) \otimes I_{d_X})\Gamma_w^T)$. Note that $(M_{MS} - M_{SS})$ is the matrix with entry $(i,j)$ given by $\trace(\Sigma_x^{-1} \sum_{\ell=0}^{\min\curly{i,j}-2} A^{\ell}B_w B_w^\top (A^\ell)^\top (A^{\abs{j-i}})^\top)$. 
Let $v_\ell = \mathsf{vec} \paren{\Sigma_x^{-1/2}  A^\ell B_w}$. Then the aforementioned matrix is equal to the gram matrix defined by 
\begin{align*}
    \bmat{0 \\ v_0^\top \\ v_1^\top \\ \vdots \\ v_{H-1}^\top} \! \! \bmat{0 \\ v_0^\top \\ v_1^\top \\ \vdots \\ v_{H-1}^\top}^\top \!\!\!+\! \bmat{0 \\ 0 \\v_0^\top \\ \vdots \\ v_{H-2}^\top} \! \! \bmat{0 \\ 0 \\v_0^\top \\ \vdots \\ v_{H-2}^\top}^\top \!\!\!+ \! \dots + \! \bmat{ 0 \\ 0 \\ \vdots \\ 0 \\ v_{0}^\top } \! \!\bmat{ 0 \\ 0 \\ \vdots \\ 0 \\ v_{0}^\top }^\top\!\!\!,
\end{align*}
and is therefore positive semidefinite. As a result,
we see that
a multi-step predictor is less efficient than a single-step predictor, and that the efficiency gap grows linearly with the prediction horizon $H$. This scaling quantitatively captures that the direct multi-step predictor has a number of parameters which scales with $H$. 

To better understand the role of system stability in determining this efficiency gap, we consider the special case of a scalar system without inputs.  Here the difference between the statistical efficiency of the single-step predictor and the multi-step predictor is characterized by the difference between the matrices
\begin{align*}
    M_{SS} =  \bmat{1 & a & a^2 & \dots &a^{H-1} \\ a & a^2 & a^3 &\dots &a^{H} \\ \vdots & & \ddots \\ 
    a^{H-1} & &\dots && a^{2(H-1)}}
\end{align*}
and
\begin{align*}
    M_{MS} = \bmat{1 & a & a^2 & \dots &a^{H-1} \\ a & 1 & a &\dots &a^{H-2} \\ \vdots & & \ddots \\ 
    a^{H-1} & &\dots && 1},
\end{align*}
from which we conclude that the difference between the two diminishes as $\abs{a} \to 1$, i.e. as the system approaches marginal stability. 
 


\section{Misspecified Setting}

We again consider a single-step predictor and a multi-step predictor applied to the measurement and sequence of future inputs. However, we now examine the general case in which measurements do not provide full state information by reincorporating partial observations, as specified by $C \in \R^{\dy \times \dx}$ and $D_v D_v^\top \succ 0$, into the dynamics~\eqref{eq: dynamics}. Due to the Markovian assumption made in fitting the predictor, this represents a misspecified setting. To ease notational burden we restrict attention to the setting without inputs and set $B=0$, although our analysis can be extended naturally to the $B \neq 0$ case.

\begin{figure}
    \centering
    \includegraphics[width=\linewidth]{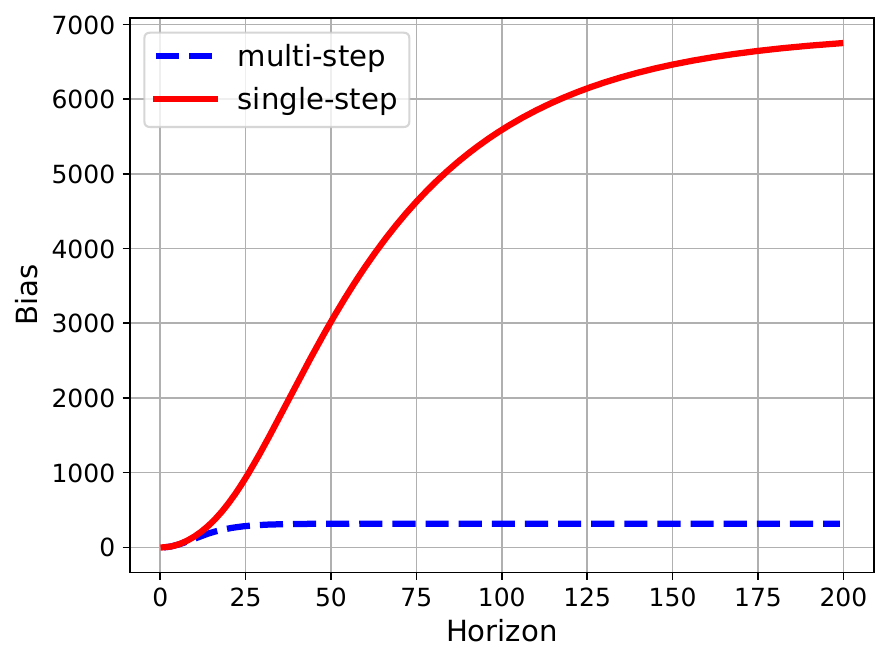}
    \caption{Comparison of the bias from the single and multi-step estimators across different horizons for the system defined in Example~\ref{example}.}
    \label{fig: bias comparison}
\end{figure}

In this setting, the loss is given by
\begin{align}
    \label{eq: misspecified prediction error}
    L(\hat G) = \bar \E \norm{y_{t+1:t+H} - \hat G y_t }^2.
\end{align}
We rewrite the dynamics in innovations form
\begin{align*}
    \hat x_{t+1} &= (A-KC) \hat x_t + K y_t = A \hat x_t + K D_e e_t \\
    y_t &= C \hat x_t + D_e e_t, 
\end{align*}
where $e_t$ is standard normal noise that is independent across time, $K$ is the Kalman gain defined as $K = ASC^\top (C S C^\top + R)^{-1}$, $S$ is the stabilizing solution to the Riccati equation defined by $A$, $C$, $D_w D_w^\top$ and $D_v D_v^\top$, and $D_e = (C S C^\top + D_v D_v^\top)^{1/2}$.  Then
\begin{align*}
    y_{t+1:t+H} = \Phi \hat x_t + G^\star y_t + \Gamma_e e_{t+1:t+H},
\end{align*}
where 
\begin{align*}
    \Phi &= \bmat{C (A-KC) \\ C A (A-KC) \\ \vdots \\ CA^{H-1} (A-KC)}, G^\star = \bmat{CK \\ C A K \\ \vdots \\ CA^{H-1} K}, \\
    \Gamma_e &= \bmat{D_e & 0 & \dots & 0\\ C K D_e & D_e & \dots \\ \vdots && \ddots \\ CA^{H-2} K D_e & \dots & CK D_e & D_e}. 
\end{align*}
Under these definitions, and exploiting that innovations are independent across time, the error~\eqref{eq: misspecified prediction error} is given by 
\begin{align*}
    L(\hat G) &= \bar \E\norm{\Phi \hat x_t + (G^\star - \hat G) y_t + \Gamma_e e_{t+1:t+H}}^2 \\
    &= \bar \E\norm{\Phi \hat x_t + (G^\star - \hat G) y_t}^2 + \norm{\Gamma_e}_F^2.
\end{align*}
Expanding $y_t = C \hat x_t + D_e e_t$,\\

\begin{align*}
    &L(\hat G) \\
    &=\! \norm{(\Phi  \!+\! ( G^\star - \hat G)C)\Sigma_{\hat x}^{1/2}}_F^2 + \norm{(G^\star - \hat G) D_e}_F^2  \!+\! \norm{\Gamma_e}_F^2, 
\end{align*}
where $\Sigma_{\hat x}$ is the stationary covariance of $\hat x_t$.  
When $\hat G$, or equivalently $\hat G_N$, is learned on the dataset of size $N$, we can decompose this quantity into an irreducible component, and a component which decays to zero as the amount of data $N\to \infty$. Denoting the irreducible component by $B(\hat G_N) \triangleq \lim_{N\to\infty} \E L(\hat G_N)$ and the reducible component by $\varepsilon(\hat G_N) \triangleq L(\hat G_N) - B(\hat G_N)$, we decompose
\begin{align*}
    L(\hat G_N) = B(\hat G_N) + \varepsilon(\hat G_N). 
\end{align*}
Unlike the well-specified setting, the irreducible component $B(\hat G_N)$ differs depending on whether we fit a single-step model or direct multi-step model. We therefore focus on comparing these bias terms rather than the rate of convergence, since this captures the fundamental difference between the two models. See \Cref{proof: multi-step misspecified} (multi-step) and \Cref{proof: single-step misspecified} (single-step) for characterizations of the rate of decay of the reducible errors $\lim_{N \rightarrow \infty} N\E[\varepsilon(\hat G_N) ]$. 

\begin{figure*}
    \centering
    \includegraphics[width=0.32\linewidth]{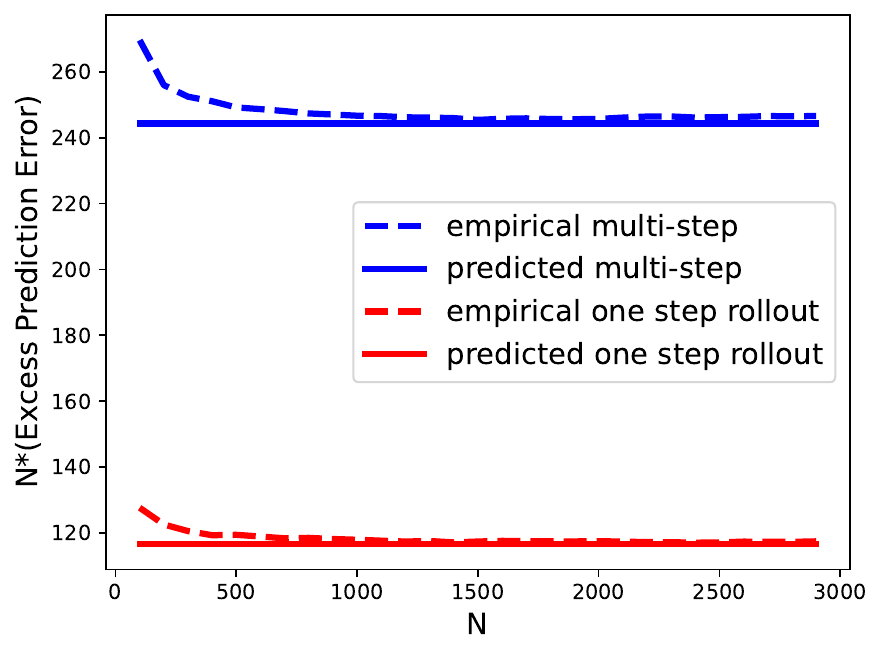}
    \includegraphics[width=0.32\linewidth]{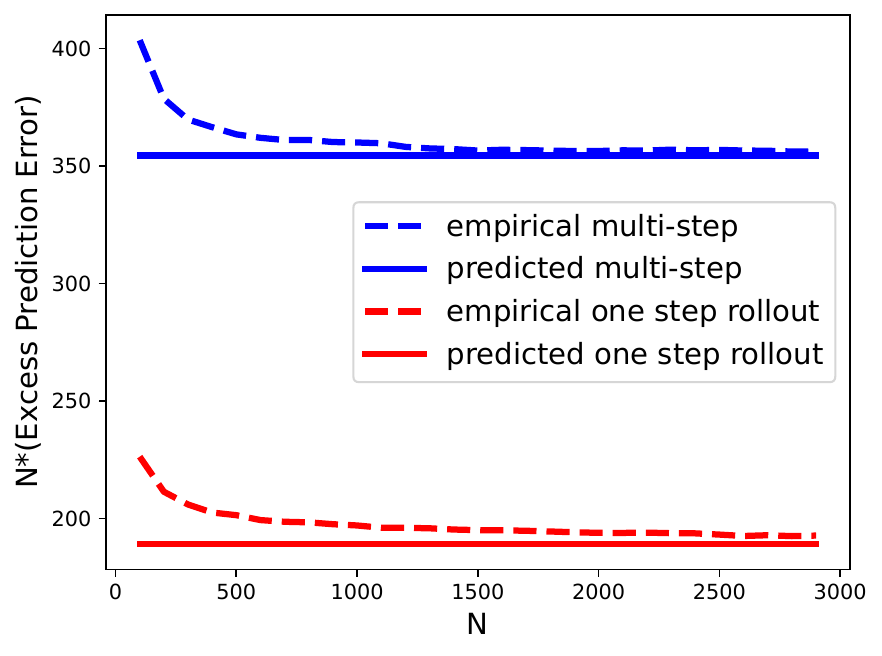}
    \includegraphics[width=0.32\linewidth]{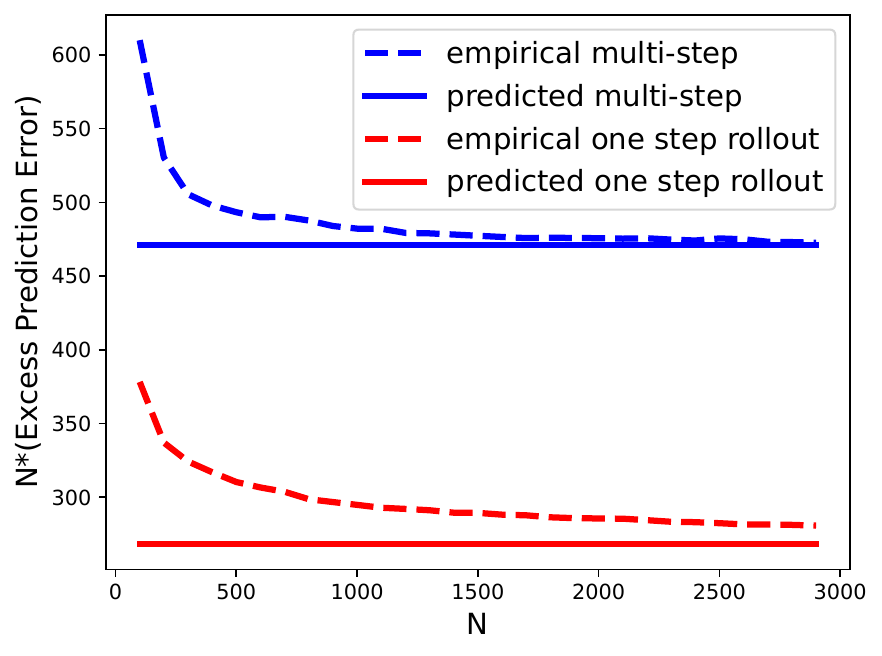}
    \caption{Convergence of $N\E[L(\hat f_H)]$ to the reducible prediction errors given in \Cref{prop: well specified multistep} (multi-step predictor) and \Cref{prop: single-step well specified} (single-step rollout) for the system defined by \Cref{eq: numerical ex system} with $a = 0.5, 0.75,0.9$ (left to right) and horizon $H=5$.}
    \label{fig: well specified sweep a}
\end{figure*}


\begin{figure*}
    \centering
    \includegraphics[width=0.32\linewidth]{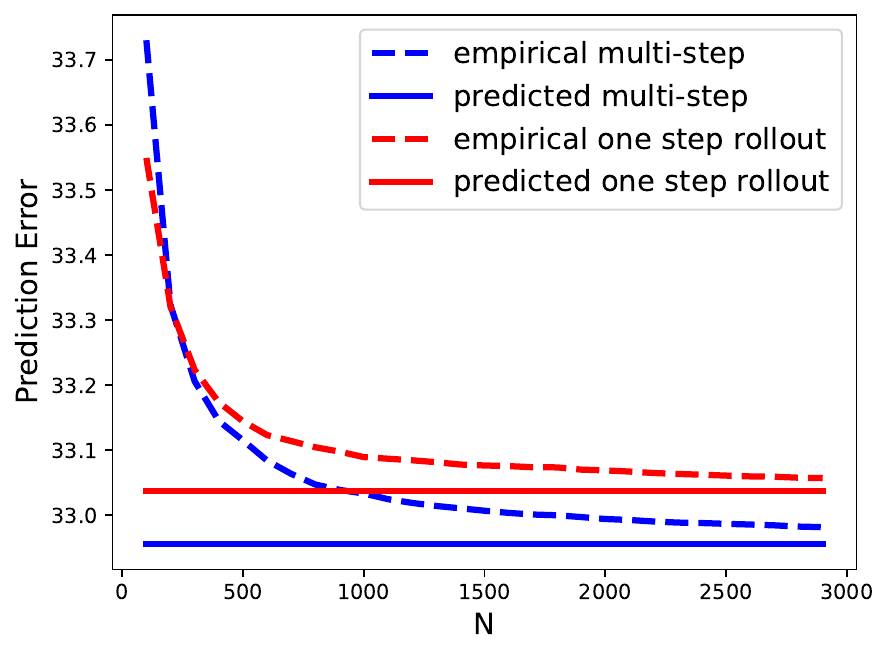}
    \includegraphics[width=0.32\linewidth]{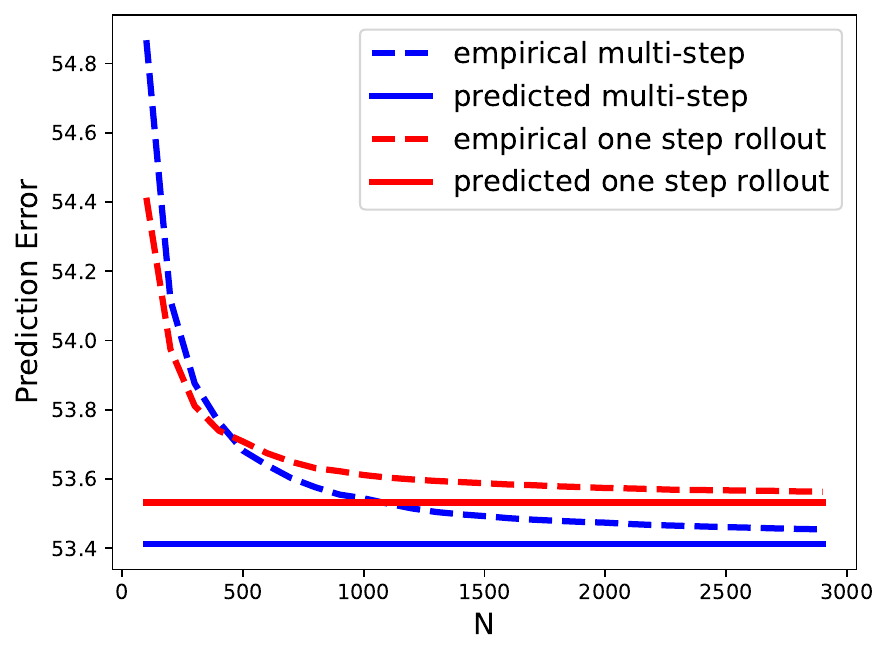}
    \includegraphics[width=0.32\linewidth]{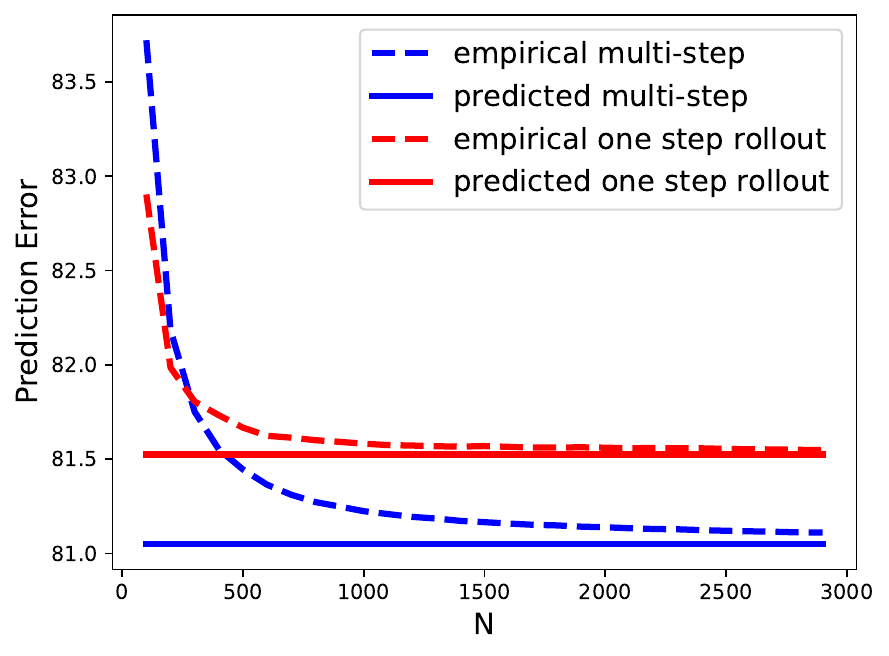}
      \caption{Convergence of $\E[L(\hat f_H)]$ to the irreducible prediction errors given in \Cref{prop: multi-step misspecified} (multi-step predictor) and \Cref{prop: single-step misspecified} (single-step rollout) for the system defined by \Cref{eq: numerical ex system} with $a = 0.5, 0.75,0.9$ (left to right) and horizon $H=5$.}
      \vspace{-16pt}
    \label{fig: misspecified sweep a}
\end{figure*}

The irreducible error for the multi-step predictor is characterized in the following proposition.
\begin{proposition}
    \label{prop: multi-step misspecified}
    The irreducible error for the multi-step predictor $\hat G_N^{MS}$ is given by 
    \begin{align*}
         B(\hat G_N^{MS}) \!=\! \trace(\Phi (\Sigma_{\hat x} - \Sigma_{\hat x} C^\top \Sigma_y^{-1} C \Sigma_{\hat x}) \Phi^\top) \! +\! \norm{\Gamma_e}_F^2. 
    \end{align*}
\end{proposition}
For the single-step predictor, the irreducible error is characterized as follows.
\begin{proposition}
    \label{prop: single-step misspecified}
    The irreducible error for the single-step predictor $\hat G_N^{SS}$ is given by  
    \begin{align*}
        B(\hat G_N^{SS}) &= \trace((\Phi + MC) \Sigma_{\hat x} (\Phi + MC)^\top) \\
        &+ \trace(M D_e D_e^\top M^\top) + \norm{\Gamma_e}_F^2, 
    \end{align*}
    where 
    \begin{align}
        \label{eq: multi-step bias characterization}
        M = G^\star - \bmat{ C A \Sigma_x C^\top \Sigma_y^{-1} \\ \vdots \\ (C A \Sigma_x C^\top \Sigma_y^{-1})^{H} },
    \end{align}
    and $\Sigma_x$ is the stationary covariance of $x_t$. 
\end{proposition}




\subsection{Comparison}

In contrast to the well-specified setting, the dominant discrepancy between the two predictors in the presence of misspecification is the bias term. Note that the bias from the multi-step predictor given in \Cref{prop: multi-step misspecified} is equal to 
\begin{align*}
    \min_{M} \trace((\Phi + MC) \Sigma_{\hat x} (\Phi + MC)^\top) + \trace(M D_e D_e^\top M^\top), 
\end{align*}
and therefore, the irreducible error for the direct multi-step never exceeds that of its single-step counterpart.

Due to the dependence of $M$  on powers of $C A \Sigma_x C^\top \Sigma_y^{-1}$ in the single-step model's bias, the spectral radius of this quantity dictates how the bias scales with the horizon. 
\Cref{lemma: spectral radius of CK+KB} shows  that the quantity $C A \Sigma_x C^\top \Sigma_y^{-1}$ which the estimate $\hat G_y$ converges to satisfies $\rho(C A \Sigma_x C^\top \Sigma_y^{-1}) \leq 1$ if $A$ is stable.  Despite this, $C A \Sigma_x C^\top \Sigma_y^{-1}$ can feature a spectral radius much larger than that of $A$, as demonstrated in the following example.  
\begin{example}
    \label{example}
    Consider the system defined by
    \begin{align*}
        A = \bmat{0.9 & 1.0 \\ 0.0 & 0.9}, \Sigma_w = I_2, C = \bmat{1, 0}, \Sigma_v = 1.
    \end{align*}
    We find that $\rho(C A \Sigma_x C^\top \Sigma_y^{-1}) = 0.99$, though $\rho(A) = 0.9$. 
\end{example}
As a consequence of this fact, the gap in bias between the multi-step error and the single-step error can grow with the horizon for moderate $H$. This is illustrated for the above example in \Cref{fig: bias comparison}.


\section{Numerical Experiments}

To validate the bounds presented in the previous sections, we consider the system defined by 
\begin{align} \label{eq: numerical ex system}
    A = \bmat{a & 1.0 \\ 0.0 & 0.75}, \Sigma_w = I_2
\end{align}
with $B = \bmat{0 & 1}^\top, C = I_2, \Sigma_v = 0$
in the well-specified setting
and, alternatively, $
    B = 0, C = [1,0], \Sigma_v = 1$
in the misspecified setting. 


\begin{figure}
    \centering
    \begin{subfigure}{0.7\linewidth}
        \centering
        \includegraphics[width=\linewidth]{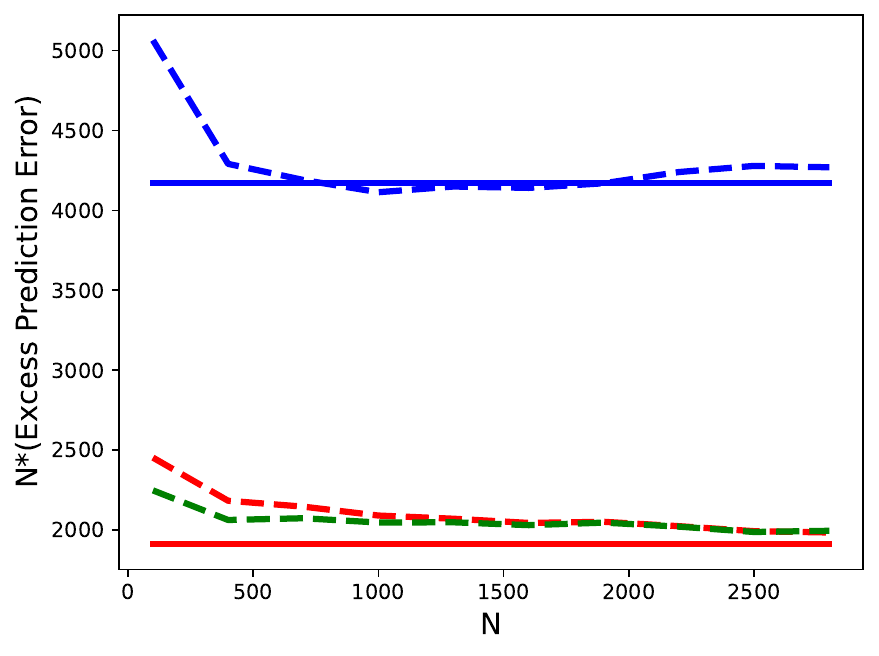}
        \caption{ Well-specified case}
        \label{fig:well-specified}
    \end{subfigure}
    \vspace{-0.5em}
    \begin{subfigure}{0.7\linewidth}
        \centering
        \includegraphics[width=\linewidth]{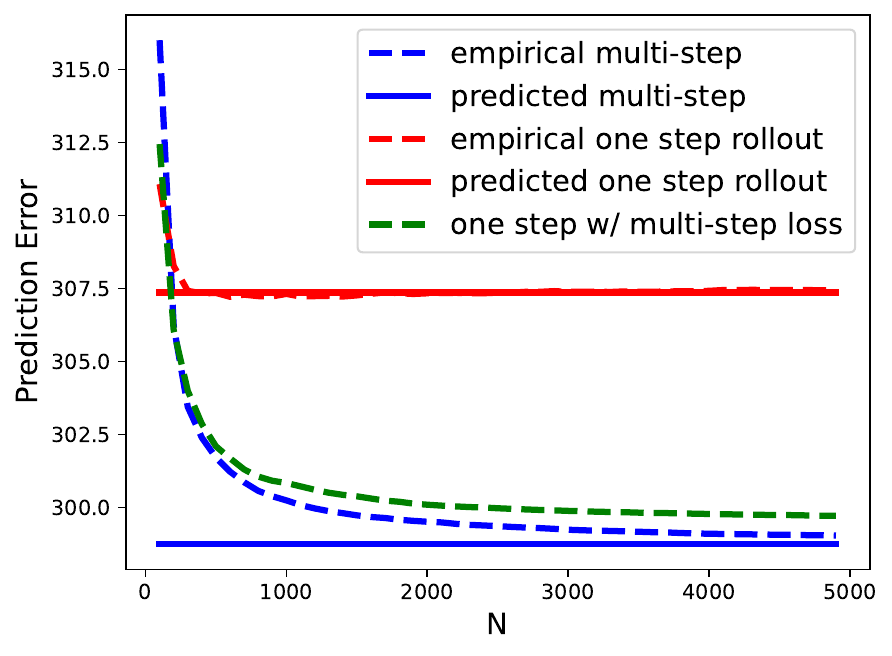}
        \caption{Misspecified case}
        \label{fig:misspecified}
    \end{subfigure}
    
    \caption{Comparison of the rate of decay for the error in the well-specified case (a) and the total  error in the misspecified case (b) for the direct multi-step predictor, and the single-step predictor trained with a single-step loss or a multi-step loss.}
    \label{fig: horizon 10}
\end{figure}

\begin{figure}
    \centering
    \begin{subfigure}{0.7\linewidth}
        \centering
        \includegraphics[width=\linewidth]{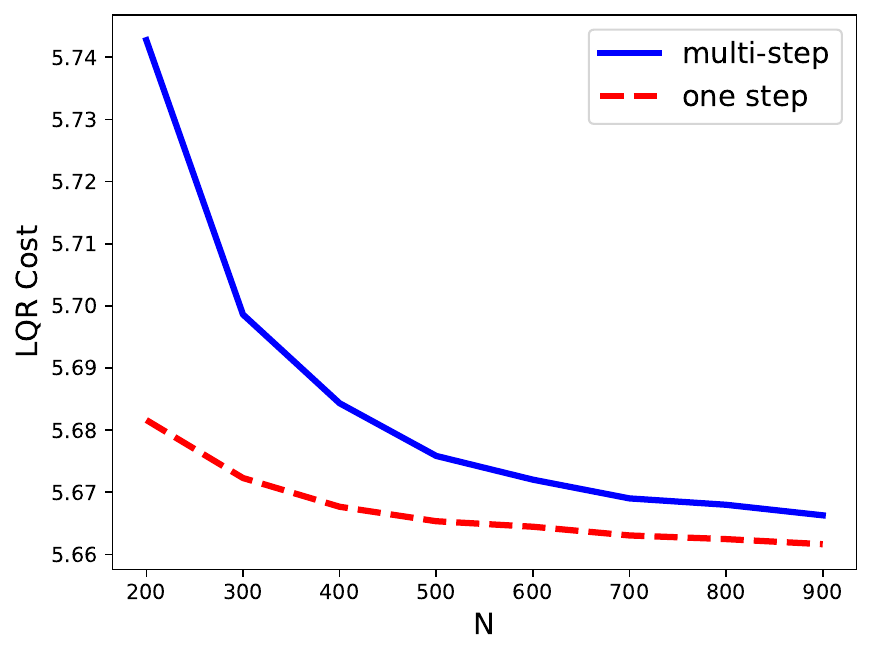}
        \caption{Infinite-horizon LQR cost, well-specified case}
        \label{fig:well-specified}
    \end{subfigure}
    \vspace{-0.5em}
    \begin{subfigure}{0.7\linewidth}
        \centering
        \includegraphics[width=\linewidth]{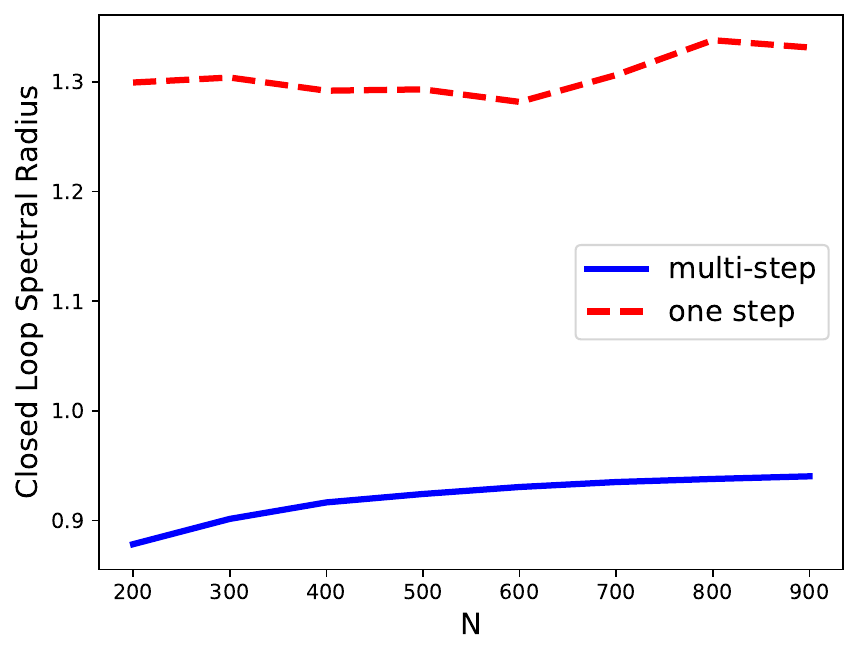}
        \caption{Spectral radius of the closed loop system, misspecified case}
        \label{fig:misspecified}
    \end{subfigure}
    
    \caption{Infinite-horizon LQR performance in the well-specified case (a) and the misspecified case (b). In (b), closed loop spectral radius greater than 1 for the one step predictor implies infinite LQR cost. }

    \label{fig: LQRcost}
\end{figure}

\textbf{Well-specified: } \Cref{fig: well specified sweep a} illustrates the well-specified setting. Specifically, we estimate $N\E[L(\hat f_H)]$ by averaging over $30,000$ datasets $D_N$ for $N \in \{ 1,...,3000\}$ to demonstrate convergence to the reducible errors given in \Cref{prop: well specified multistep} and \Cref{prop: single-step well specified}. In these figures, we fix $H=5$ and vary $a$ across $0.5, 0.75$, and $0.9$. 

\textbf{Misspecified: }  \Cref{fig: misspecified sweep a} illustrates the misspecified setting. Specifically, we estimate $\E[L(\hat f_H)]$ by averaging over $1,000$ datasets $D_N$ for $N \in \{ 1,...,3000\}$ to demonstrate convergence to the irreducible errors given in \Cref{prop: multi-step misspecified} and \Cref{prop: single-step misspecified}. In these figures, we fix $H=5$ and vary $a$ across $0.5, 0.75$, and $0.9$. 

\textbf{Multi-step loss: } In \Cref{fig: horizon 10}, we compare the multi-step predictor with the single-step predictor trained using a single-step loss, and a multi-step loss \eqref{eq: multistep loss} with $a=0.9$ and $H=10$. We see that in the well-specified setting, the rate of decay for the prediction error of the single-step model trained with a multi-step loss matches the rate of decay for the prediction error using a single-step loss. However, in the presence of misspecification, the prediction error converges nearly to the level of the direct multi-step predictor. The function class for the predictor \eqref{eq: multistep loss} strictly less expressive than the direct multi-step predictor, \eqref{eq: multistep LS}, which explains why the direct multi-step loss still incurs less bias. For the single-step model with a multi-step loss, $\hat G$ is fit using gradient descent initialized from the single-step predictor fit with a single-step loss, and using a step size $2e-5$.  

\textbf{Control Performance: } In \Cref{fig: LQRcost}, we consider a closed-loop control setting in which the control inputs are selected using predictions from either single-step or multi-step models, each trained on datasets of size $N$. In particular the control input is selected via model predictive control using a horizon $H=20$ with stage costs $c(y_t, u_t) = \norm{y_t}^2 + \norm{u_t}^2$ and $y_{t+H}$ constrained to 0.  Panel (a) shows the infinite-horizon LQR cost, $\lim_{T \to \infty} \E\Bigl[\sum_{t=1}^T y_t^\top y_t + u_t^\top u_t\Bigl]$ incurred by this controller in the well-specified setting for a system with $a = 0.9$. averaged over 1,000 datasets. The multi-step predictor uses the same horizon, $H=5$ as the MPC horizon. In the low-data regime, single-step rollouts result in lower cost than multi-step prediction.

Panel (b) shows the spectral radius of the resulting closed-loop system in the misspecified setting for the same system ($a = 0.9$, $H = 20$), averaged over 1,000 datasets. A spectral radius greater than 1 for the single-step rollout indicates that the closed-loop system is unstable, and the associated infinite-horizon LQR cost diverges.

\section{Conclusion}
In this work, we present a novel theoretical comparison of the asymptotic prediction error associated with autoregressive rollouts of single-step predictors and direct multi-step predictors. Our analysis offers insight into when each modeling approach is preferable. Specifically, we show that for well-specified model classes, autoregressive rollouts of single-step predictors achieve lower asymptotic prediction error. However, in the presence of model misspecification due to an incorrect Markovian assumption, multi-step predictors can significantly outperform their single-step counterparts.

These findings provide a foundation for more informed model design in learning-based control and forecasting. Promising directions for future work include: (1) developing a rigorous analysis of intermediate approaches, such as the single-step model trained with a multi-step loss, which we investigate only empirically in this work; and (2) extending our analysis beyond the white-noise input assumption to study how each of these prediction approaches performs in a closed-loop control setting.

\section*{Acknowledgements}
This work is supported in part by NSF Award SLES-2331880, NSF CAREER award ECCS-2045834, and AFOSR Award FA9550-24-1-0102.

\bibliographystyle{IEEEtranN}
\bibliography{refs}

\appendices
\section{}
\begin{lemma}
\label{lemma: spectral radius of CK+KB}
    Assume $\rho(A) < 1$. Then, 
    \begin{align*}
        \rho(C A \Sigma_x C^\top \Sigma_y^{-1}) \leq 1. 
    \end{align*}
\end{lemma}
\begin{proof}[Proof of \Cref{lemma: spectral radius of CK+KB}]
Note that $ C A \Sigma_{ X} C^\top \Sigma_y^{-1} = \bar \E[y_{t+1} y_t^\top] \bar \E[y_t y_t^\top]^{-1}$. From the stationarity of the process,
\begin{align*}
    \bar \E \bmat{y_t \\ y_{t+1}} \bmat{y_t \\ y_{t+1}}^\top = \bmat{\Sigma_y & \Sigma_{y+} \\ \Sigma_{y+} & \Sigma_y}. 
\end{align*}
By a Schur complement,  $\Sigma_y - \Sigma_{y+} \Sigma_y^{-1} \Sigma_{y+} \succeq 0,$ or $(\Sigma_y^{-1/2} \Sigma_{y+} \Sigma_{Y}^{-1/2})^2 \preceq I$. Then $\norm{\Sigma_y^{-1/2} \Sigma_{y+} \Sigma_{Y}^{-1/2}} \leq 1$. For $i=1,\dots, \dy$, it holds that $\abs{\lambda_i(\Sigma_{y+} \Sigma_{Y}^{-1})} =\abs{\lambda_i(\Sigma_y^{-1/2} \Sigma_{y+} \Sigma_{Y}^{-1/2})} \leq 1$, and thus $\rho(\Sigma_{y+} \Sigma_y^{-1}) \leq 1$. 
\end{proof}

For the proofs of the four main propositions in the main paper, we use the following facts. It holds by the Birkoff-Khinchin theorem that $\lim_{N\to\infty} \frac{1}{N}\sum_{t=1}^{N} z_tz_t^\top \to \Sigma_z$. By Slutky's theorem, if $\lim_{N} X_N \overset{d}{=} X$ and $\lim_N Y_N = c$ for $X$ a random variable and $c$ a constant, then $\lim_N X_n Y_n \overset{d}{=} X c$. It holds by the dominated convergence theorem (DCT) that $\lim_N \E \norm{(\hat G_N - G^\star) \Sigma_{z}^{1/2}}_F^2 = \E\brac{\lim_{N} \norm{(\hat G_N - G^\star) \Sigma_{z}^{1/2}}_F^2}$.

We use these facts throughout the proofs.

\subsection{Proof of \Cref{prop: well specified multistep}}

\begin{proof}
    For ease of notation, we will refer to $\hat G^{MS}_N$ as $\hat G_N$. By definition of the least squares solution,  
\begin{align*}
    \hat G_N - G^\star &= \sum_{t=1}^{T-H+1} \Gamma_w w_{t+1:t+H-1} z_t^\top \Bigg(\sum_{t=1}^{T-H+1} z_tz_t^\top\Bigg)^{-1}.
\end{align*}
By Slutsky and DCT,
\begin{align*}
    &\lim_{N\to\infty} N \E \norm{(\hat G_N - G^\star) \Sigma_z^{1/2}}_F^2 \\&= \lim_{N\to\infty} N \E  \norm{ \sum_{t=1}^{T-H+1} \Gamma_w w_{t+1:t+H-1} z_t^\top \Sigma_z^{-1/2}}_F^2.
\end{align*}
Expanding the Frobenius norm results in the trace
\begin{align*}
\sum_{t,k=1}^{T-H+1} \trace\paren{ \Gamma_w w_{t+1:t+H-1} z_t^\top \Sigma_z^{-1} z_k w_{k+1:k+H-1}^\top \Gamma_w^\top}. 
\end{align*}
Consider the expectation of the summand for any pair of indices $t$ and $k$. If $t = k$, we may use the fact that $w_{t+1:t+H-1}$ is independent from $z_t$ to conclude
\begin{align*}
    &\E\brac{\trace\paren{ \Gamma_w w_{t+1:t+H-1} z_t^\top \Sigma_z^{-1} z_k w_{k+1:k+H-1}^\top \Gamma_w^\top}} \\ 
    &= \trace\paren{\Gamma_w\E \brac{w_{t+1:t+H-1} \E \brac{z_t^\top \Sigma_z^{-1} z_t} w_{t+1:t+H-1}^\top} \Gamma_w^\top} \\
    &= \trace\paren{\Gamma_w (\dx + H \du) I \Gamma_w}.
\end{align*}
If instead $t = k +m$ for $m > 0$, we have the following term 
\begin{align*}
    \E\brac{\trace\paren{ \Gamma_w w_{k+m+1:k+m+H-1} z_{k+m}^\top \Sigma_z^{-1} z_k w_{k+1:k+H-1}^\top \Gamma_w^\top}}.
\end{align*}
Expanding $z_{k+m}$ provides
\[
    z_{k+m} = \bmat{A^m x_k + \sum_{\ell=0}^{m-1} A^{m-1-\ell} (B u_{k+\ell} + B_w w_{k+\ell}) \\ u_{k+m: k+m+H-1}},
\]
and thus the expectation simplifies to
\begin{align*}
     &\E\brac{\trace\paren{ \Gamma_w w_{t+1:t+H-1} z_t^\top \Sigma_z^{-1} z_k w_{k+1:k+H-1}^\top \Gamma_w^\top}} \\ 
    &= \trace\paren{\Gamma_w \trace(A^m) L^m \Gamma_w^\top},
\end{align*}
where $L$ is the downshift matrix. Summing over all indices thus leads to the characterization in the statement.
\end{proof}

\subsection{Proof of \Cref{prop: single-step well specified}}
\begin{proof}

For ease of notation, we will refer to $\hat G^{SS}_N$ as $\hat G_N$. It holds that
\begin{align*}
    \hat G_N - G &= \Gamma (I_H \otimes (\bmat{\hat G_y - G_y & \hat G_u - G_u}  ))F \\
    &\quad \quad  + {\cal O}((\bmat{\hat G_y -G_y& \hat G_u - G_u} )^2)
\end{align*}
where 
\begin{align*}
        F = \bmat{I_{\dx} \\ &I_{\du} \\A & B \\ & & I_{\du} \\ 
        \vdots \\ A^{H-1} & A^{H-2}B & &\dots  & & B \\ & & & & & & I_{\du}}
    \end{align*}
    and
    \begin{align*}
        \Gamma = \bmat{I_{\dx} \\ A & I_{\dx} \\ \vdots \\ A^{H-1} & A^{H-2} & \dots & I_{\dx}}.
    \end{align*}

Note that 
\begin{align*}
    &\bmat{\hat G_y - G_y& \hat G_u - G_u} \\
    &= \sum_{t=1}^{N-1}B_w w_t\bmat{x_t \\ u_t} \Bigl(\sum_{t=1}^{N-1} \bmat{x_t \\ u_t} \bmat{x_t \\ u_t}^\top \Bigl)^{-1}.
\end{align*} Birkoff-Khinchin combined with Slutsky's theorem shows that this quantity scales as $\calO(1/N)$. Thus, in the limit, higher order terms vanish and 
\begin{align*}
    &\lim_{N\to\infty} N \E \brac{\norm{(\hat G_N - G^\star) \Sigma_z^{1/2}}_F^2} \\
    &=\! \lim_{N\to\infty}\!\! N \E \brac{\norm{\Gamma (I_H \!\otimes\! (\bmat{\hat G_y \!-\! G_y \!&\!\! \hat G_u \!- \!G_u} \! )\!)F \Sigma_z^{1/2}}_F^2}.
\end{align*}
Vectorizing the content of the above Frobenius norm leads to the expression
\begin{align*}
    \lim_{N\to\infty} \!\!N\E \brac{\norm{(\Sigma_z^{1/2}F^T \!\otimes \!\Gamma)L\VEC (\bmat{\hat G_y \!-\! G_y\! & \!\!\hat G_u \!-\! G_u}  )}^2} 
\end{align*}
where $L = \sum_{i = 1}^H e_i \otimes I_{\dx + \du} \otimes e_i \otimes I_{\dx}$ and $e_i$ is the $i$th column of $I_H$. 
Note that 
\begin{align*}
    &\VEC (\bmat{\hat G_y \!-\! G_y\! & \!\!\hat G_u \!-\! G_u}) \\
    &= \Bigl(\Bigl(\sum_{t=1}^{N-1} \bmat{x_t \\ u_t} \bmat{x_t \\ u_t}^\top \Bigl)^{-1} \otimes B_w \Bigl)\Bigl(\sum_{t=1}^{N-1} (\bmat{x_t \\ u_t} \otimes I_{d_x})w_t\Bigl).
\end{align*}
A combination of Birkoff-Khinchin and Slutsky's shows that 
\begin{align*}
    &N\VEC (\bmat{\hat G_y \!-\! G_y\! & \!\!\hat G_u \!-\! G_u})^\top \VEC (\bmat{\hat G_y \!-\! G_y\! & \!\!\hat G_u \!-\! G_u}) \\
    &\rightarrow \Sigma_{x,u}^{-1} \otimes B_w B_w^\top 
\end{align*}
where $\Sigma_{x,u}$ is the stationary covariance of $\bmat{x_t^\top & u_t^\top}^\top$. With this, the above expression becomes  
\begin{align*}
    \trace ((\Sigma_{x,u}^{-1} \otimes B_wB_w^\top)L^T(F\Sigma_zF^T \otimes \Gamma^T\Gamma)L).
\end{align*}
Simplifying concludes the proof.
\end{proof}
\subsection{Proof of \Cref{prop: multi-step misspecified}}
\label{proof: multi-step misspecified}

\begin{proof}

For ease of notation, we will refer to $\hat G^{MS}_N$ as $\hat G_N$. The least squares identification error is 
\begin{align*}
    \hat G_N\! -\! G^\star \!=\! \sum_{t=1}^{N-H} (\Phi \hat x_t + \Gamma_e e_{t+1:t+H}) y_t^\top \paren{\sum_{t=1}^{N-H} y_t y_t^\top}^{-1}.
\end{align*}
By expanding $y_t = C \hat x_t + D_e e_t$, and replacing the sample average covariance of $y_t$ with its population counterpart this becomes 
\begin{align*}
    &\frac{1}{N-H}\sum_{t=1}^{N-H} \Phi \hat x_t \hat x_t^\top C^\top \Sigma_y^{-1} +\\
    &\frac{1}{N-H}\sum_{t=1}^{N-H} \Phi \hat x_t e_t^\top D_e^\top \Sigma_y^{-1} \!+\! \sum_{t=1}^{N-H}  \Gamma_e e_{t+1:t+H}  y_t^\top \Sigma_y^{-1}.
\end{align*}
The first term converges to a bias $\Phi \Sigma_{\hat x} C^\top \Sigma_y^{-1}$. Denote the other term $\tilde E$.
It holds by DCT and Slutky's Theorem that 
\begin{align*}
    &\E \brac{\norm{(\Phi + G^\star - \hat G_N) C \Sigma_{\hat x}^{1/2}}_F^2} - \E \brac{\norm{\tilde E C \Sigma_{\hat x}^{1/2}}_F^2} \\
    &\to \norm{\Phi \Sigma_{\hat x}^{1/2}(I - \Sigma_{\hat x}^{1/2} C^\top \Sigma_y^{-1} C \Sigma_{\hat x}^{1/2})}_F^2 \\
    &\E \brac{\norm{(G^\star - \hat G_N) D_e}_F^2} - \E \brac{\norm{\tilde E D_e}_F^2} \\
    &\to \norm{\Phi \Sigma_{\hat x} C^\top \Sigma_y^{-1} D_e}_F^2.
\end{align*}
Then the overall loss becomes 
\begin{align*}
    L(\hat f_H) = \trace(\Phi (\Sigma_{\hat x} - \Sigma_{\hat x} C^\top \Sigma_y^{-1} C \Sigma_{\hat x}) \Phi^\top) + \varepsilon_N
\end{align*}
where $\lim_{N \to \infty} N \E \varepsilon_N= \lim_{N \rightarrow \infty} 1/N \E \trace\paren{\tilde E \Sigma_y \tilde E}$. It holds that
\begin{align*}
    &(N-H)^2 \E \trace\paren{\tilde E \Sigma_y \tilde E} =\\ &\E\norm{\sum_{t=1}^{N-H} ( \Phi \hat x_t e_t^\top D_e^\top + \Gamma_e e_{t+1:t+H} y_t^\top) \Sigma_y^{-1/2}}_F^2.
\end{align*}
We separately study the quantities 
\begin{align*}
    \sum_{t=1}^{N-H} \Phi \hat x_t e_t^\top D_e^\top \Sigma_y^{-1/2}
\mbox{ and }
     \sum_{t=1}^{N-H} \Gamma_e e_{t+1:t+H} y_t^\top \Sigma_y^{-1/2}
\end{align*}
along with their cross terms. It follows from the derivations of \Cref{prop: well specified multistep} that the expected Frobenius norm of the second term is asymptotically characterized by $\trace(\Gamma_e (M_1 \otimes I_{\dy}) \Gamma_e^\top)$ where 
\begin{align*}
        M_1 &\!=\!\!\bmat{\trace(I) & \trace(\Sigma_y^{(1)}) & \dots & &\trace(\Sigma_y^{(H-1)}) \\ 
        \trace(\Sigma_y^{(1)}) & \trace(I) & \trace(\Sigma_y^{(1)}) & \!\!\dots\!\! & \!\!\trace(\Sigma_y^{(H-2)}) \!\!\\ \vdots & & &\!\!\!\ddots\!\!\! \\ \trace(\Sigma_y^{(H-1)})\!\!\! & \dots & & &\trace(I)
        }, \\
        \Sigma_y^{(i)} &= (C A^{i} \Sigma_{\hat x} C^\top + CA^{i-1} K D_e D_e^\top) \Sigma_y^{-1}. 
\end{align*} 

The expected Frobenius norm of the second term converges to $\trace(\Phi \Sigma_{\hat x} \Phi^\top) \trace(D_e^\top \Sigma_{Y}^{-1} D_e)$.
It remains to handle the cross terms:
\begin{align*}
   2 \E \sum_{t=1}^{N-H} \sum_{k=1}^{N-H} \trace\paren{\Phi \hat x_t e_t^\top D_e^\top \Sigma_y^{-1} y_k e_{k+1:k+H}^\top \Gamma_e^\top}.
\end{align*}
The terms of this sum with $t \notin \curly{k+1, \dots, k+H}$ are zero. For the nonzero terms we may express $\hat x_t = A^{t-k} \hat x_k + \sum_{\ell=0}^{t-k-1} A^{t-k-1-\ell} K D_e e_{k+\ell}$. The above term 
simplifies to 
\begin{align*}
   & 2 (N-H)   \sum_{t=k+1}^{\min\curly{k+H, N-H}} \trace\Bigg(\Phi  (A^{t-k} \Sigma_{\hat x} C^\top \\
   &+ A^{t-k-1} K D_e D_e^\top )\Sigma_y^{-1} D_e (e_{t-k,H}^\top \otimes I_{\dy} )\Gamma_e^\top\Bigg).
\end{align*}
Asympototically, $N/(N-H)^2$ times this quantity converges to $2 \trace\Bigg(\Phi M_2 (I_H \otimes (A \Sigma_{\hat x} C^\top + K D_e D_e^\top) \Sigma_y^{-1} D_e )\Gamma_e^\top\Bigg),$ where $M_2 = \bmat{I &A & \dots & A^{H-1}}.$ Then the reducible error can be characterized as 
\begin{align*}
        &\lim_{N\to \infty} \!\!\! N \!\E \brac{\varepsilon_N } \\
        &=\! \trace(\Phi \Sigma_{\hat x} \Phi^\top\!)\! \trace(D_e^\top \Sigma_{Y}^{-1} D_e) \!+\! \trace\paren{\Gamma_e (M_1 \otimes I_{\dy}) \Gamma_e^\top\!} \\ 
        &+2 \trace\Bigg(\Phi M_2 (I_H \otimes (A \Sigma_{\hat x} C^\top + K D_e D_e^\top) \Sigma_y^{-1} D_e )\Gamma_e^\top\Bigg). 
    \end{align*}

\end{proof}

\subsection{Proof of \Cref{prop: single-step misspecified}}
\label{proof: single-step misspecified}
\begin{lemma}
    \label{lemma: single-step misspecified}
    Let 
    \begin{align}
    \tilde E\! =\! \frac{1}{N\!-\!1}  \sum_{t=1}^{N\!-\!1} ( (C(A-KC) \hat x_t e_t^\top D_e^\top \!+\! D_e e_{t+1} y_t^\top) \Sigma_y^{-1}. 
\end{align}
and ${\cal X,Y} \in \R^{d_Y \times d_Y}$
Then, 
\begin{align*}
    &\Omega({\cal X,Y}) \triangleq \lim_{N \to \infty} \E \Bigl[\trace\Bigl(\VEC(\tilde E)\VEC(\tilde E)^T({\cal X} \otimes {\cal Y})\Bigl)\Bigl] \\
    &= \trace \Bigl(\Sigma_y^{-1}D_eD_e^T\Sigma_y^{-1}{\cal X}\Bigl)\\
    &\quad \cdot \trace\Bigl(C(A-KC)\Sigma_{\hat x} (C(A-KC))^T \cal{Y}\Bigl ) \\
&+\trace\Bigl(\Sigma_y^{-1}D_eD_e^T J_{d_Y} \Sigma_y^{-1}{\cal X}\Bigl) \\
&\quad \cdot \trace \Bigl( (C(A-KC)A\Sigma_{\hat x}C^T + KD_eD_e^T)J_{d_Y}{\cal Y}\Bigl) \\
&+\trace\Bigl({\cal X}^\top \Sigma_y^{-1}D_eD_e^T J_{d_Y} \Sigma_y^{-1}\Bigl) \\
&\quad \cdot \trace \Bigl( {\cal Y}^\top (C(A-KC)A\Sigma_{\hat x}C^T + KD_eD_e^T)J_{d_Y}\Bigl) \\
    &+ \trace \Bigl( \Sigma_y^{-1} {\cal X}\Bigl) \trace \Bigl( D_eD_e^T {\cal Y}\Bigl) 
\end{align*}
where $J_{d_Y} \in \R^{d_Y \times d_Y}$ is the matrix with every element equal to $1$. 
\end{lemma}
\begin{proof}[Proof of Lemma]
    Note that $\VEC(\tilde E) = S_1 + S_2$ where 
\begin{align*}
    S_1 = \Bigl(\Sigma_y^{-1} \otimes I_{d_Y}\Bigl)\sum_{t=1}^N(D_e \otimes C(A-KC)\hat x_t)e_t
\end{align*}
and 
\begin{align*}
    S_2 &= \Bigl(\Sigma_y^{-1}  \otimes I_{d_Y}\Bigl)\sum_{t=1}^N(y_t \otimes D_e)e_{t+1} \\
    &= \Bigl(\Sigma_y^{-1}  \otimes I_{d_Y}\Bigl)\sum_{t=1}^N\boldsymbol{K}^{(d_Y, d_Y)}(D_e \otimes y_t)e_{t+1}
\end{align*}
where $\boldsymbol{K}^{(d_Y, d_Y)}= \sum_{k,l=1}^{d_Y} e_{d_Y, l}e_{d_Y,k}^\top \otimes e_{d_Y, k}e_{d_Y,l}^\top$ is the commutation matrix of dimension $d_Y \times d_Y$. 
Then
\begin{align*}
    &\trace\Bigl(\VEC(\tilde E)\VEC(\tilde E)^T({\cal X} \otimes {\cal Y})\Bigl)\\
    &= \trace((S_1S_1^T + S_1S_2^T + S_2S_1^T + S_2S_2^T) ({\cal X} \otimes {\cal Y})).
\end{align*}
The result follows by noting that 
\begin{align*}
    &\lim_{N \rightarrow \infty} \E[S_1S_1^T] \\
    &\quad \quad = \Sigma_y^{-1}D_eD_e^T\Sigma_y^{-1} \otimes C(A-KC)\Sigma_{\hat x}(C(A-KC))^T, \\
    &\lim_{N \rightarrow \infty} \E[S_1S_2^T] = \sum_{k,l=1}^{d_Y} \Sigma_y^{-1}D_eD_e^Te_{d_Y, l}e_{d_Y,k}^\top \Sigma_y^{-1} \\
    &\quad \quad \quad \quad   \otimes (C(A-KC)A\Sigma_{\hat x}C^T + KD_eD_e^T)e_{d_Y, k}e_{d_Y,l}^\top,
    \end{align*}
    and 
    \begin{align*}
    &\lim_{N \rightarrow \infty} \E[S_2S_2^T] {=} \Sigma_y^{-1} \otimes D_eD_e^T. 
\end{align*}
\end{proof}

\begin{proof}[Proof of \Cref{prop: single-step misspecified}]
It holds that 
\begin{align*}
    &\hat G_y = \sum_{t=1}^{N-1} y_{t+1} y_t^\top \paren{\sum_{t=1}^{N-1} y_t y_t^\top}^{-1} \\
    &\!=\! CK \!+\! \sum_{t=1}^{N-1} (C(A\!-\!KC) \hat x_t \!+\! D_e e_{t+1})y_t^\top \!\paren{\!\sum_{t=1}^{N-1} y_t y_t^\top\!}^{-1}. 
\end{align*}
We may use Birkoff-Khinchin and Slutsky's Theorem to replace $\paren{\!\sum_{t=1}^{N-1} y_t y_t^\top\!}^{-1}$ with $\frac{1}{N-1} \Sigma_y^{-1}$. 
Expanding $y_t = C \hat x_t + D_e e_t$, this becomes 
\begin{align*}
    &CK + \frac{1}{N-1} \sum_{t=1}^{N-1} C(A-KC) \hat x_t \hat x_t^\top C^\top \Sigma_y^{-1} \\
    &+\frac{1}{N-1}  \sum_{t=1}^{N-1} ( (C(A-KC) \hat x_t e_t^\top D_e^\top + D_e e_{t+1} y_t^\top) \Sigma_y^{-1}.
\end{align*}
Using convergence of $\frac{1}{N-1}\sum_{t=1}^{N-1} \hat x_t \hat x_t^\top$ to $\Sigma_x$, we can say that  
\begin{align*}
    \hat G_y &\approx CK + C(A-KC) \Sigma_{\hat x} C^\top \Sigma_y^{-1} \\
    &+\frac{1}{N-1}  \sum_{t=1}^{N-1} ( (C(A-KC) \hat x_t e_t^\top D_e^\top + D_e e_{t+1} y_t^\top) \Sigma_y^{-1} \\
    &=  C A \Sigma_x C^\top \Sigma_y^{-1} + \tilde E \\
\end{align*}
where
$\approx$ denotes asymptotic equality in distribution and 
\begin{align}
    \tilde E \!\triangleq \!\frac{1}{N\!-\!1}  \!\sum_{t=1}^{N\!-\!1} ( (C(A-KC) \hat x_t e_t^\top D_e^\top \!+\! D_e e_{t+1} y_t^\top) \Sigma_y^{-1}. 
\end{align}
For ease of notation, we will refer to $\hat G^{SS}_N$ as $\hat G_N$. We can then show that 
\begin{align*}
    \hat{G}_N &= \bmat{\hat G_y \\ \hat G_y^2 \\ \vdots \\ \hat G_y^H} 
\approx
\bmat{C A \Sigma_x C^\top \Sigma_y^{-1} \\ (C A \Sigma_x C^\top \Sigma_y^{-1})^2 \\ \vdots \\ \paren{C A \Sigma_x C^\top \Sigma_y^{-1}}^H  } + \Gamma(I_H \otimes \tilde E)F \\
    &+ (L_H \otimes I_{d_Y})(\Gamma (I_H \otimes \tilde E))^2F + O(\tilde E ^3)
\end{align*}
where 
\begin{align*}
    \Gamma = \bmat{I_{d_Y} \\ C A \Sigma_x C^\top \Sigma_y^{-1} & I_{d_Y}\\ \vdots \\ (C A \Sigma_x C^\top \Sigma_y^{-1})^{H-1} & (C A \Sigma_x C^\top \Sigma_y^{-1})^{H-2} & \dots & I }
\end{align*}
and $F$ is the first block column of $\Gamma$ and $L_H$ is the $H \times H$ downshift matrix. For $M$ as in the proposition statement,
\begin{align*}
    G^\star - \hat G_N &\approx M - \Gamma(I_H \otimes \tilde E)F \\
    &- (L_H \otimes I_{d_Y})(\Gamma (I_H \otimes \tilde E))^2F + O(\tilde E ^3).
\end{align*}
Plugging this in to $L(\hat f_H)$, the loss can be reduced to 

\begin{align*}
    &\trace \Bigl(\Phi \Sigma_x \Phi^\top + \Phi\Sigma_{\hat x}C^\top M^\top + MC\Sigma_{\hat x}\Phi^\top + M\Sigma_yM^\top \Bigl)  \\
    &+ \norm{\Gamma_e}_F^2 + \E\Bigl[\norm{\Gamma(I_H \otimes \tilde E)F \Sigma_y^{\frac{1}{2}}}_F^2 \\
    &-\!2\trace \Bigl(\!(M\Sigma_y \!+\! \Phi\Sigma_xC^\top)\Bigl( (L_H \!\otimes\! I_{d_Y})(\Gamma (I_H \!\otimes \!\tilde E))^2F\Bigl)^\top \Bigl)\Bigl]\\
    & + {\cal O}(\tilde E ^3). 
\end{align*}
Note that
\begin{align*}
    &\E\!\norm{\Gamma(I_H \otimes \tilde E)F \Sigma_y^{\frac{1}{2}}}_F^2 \!=\! \E\norm{(\Sigma_y^{\frac{1}{2}}F^\top\! \otimes\! \Gamma)L\VEC (\tilde E)}_F^2
\end{align*}
where 
$L = \sum_{i = 1}^H e_i \otimes I_{d_Y} \otimes e_i \otimes I_{d_Y}$ and $e_i$ is the $i$th column of $I_H$. Then,
\begin{align*}
    &\E\Bigl[\norm{(\Sigma_y^{\frac{1}{2}}F^\top \otimes \Gamma)L\VEC (\tilde E)}_F^2 \\
    &= \sum_{i,j=1}^{d_Y}\trace\Bigl(\VEC(\tilde E) \VEC(\tilde E)^\top\Bigl((e_i^\top \otimes I_{d_Y})F\Sigma_yF^\top(e_j \otimes I_{d_Y}) \\
    &\otimes (e_i^\top \otimes I_{d_Y})\Gamma^\top\Gamma(e_j \otimes I_{d_Y})\Bigl)\Bigl) \\
    &=\sum_{i,j=1}^n\Omega\Bigl((e_i^\top \otimes I_{d_Y})F\Sigma_yF^\top(e_j \otimes I_{d_Y}), \\
        & \quad \quad(e_i^\top \otimes I_{d_Y})\Gamma^\top\Gamma(e_j \otimes I_{d_Y})\Bigl)
\end{align*}
where $\Omega(\cdot, \cdot)$ is defined in \Cref{lemma: single-step misspecified}.
Also note that
\begin{align*}
    &2\trace \Bigl((M\Sigma_y + \Phi\Sigma_xC^\top)\Bigl( (L_H \otimes I_{d_Y})(\Gamma (I_H \otimes \tilde E))^2F\Bigl)^\top\Bigl)\Bigl] \\
    &= 2\VEC\Bigl(\Gamma(I_H \otimes \tilde E)\Bigl)^\top\VEC\Bigl((L_H \otimes I_{d_Y})^\top\\
    &\quad \quad \cdot (M\Sigma_y + \Phi\Sigma_xC^\top)F^\top(I_H \otimes \tilde E^\top)\Gamma^\top\Bigl) \\
    &= 2\trace \Bigl(\VEC(\tilde E) \VEC(\tilde E)^\top L^\top(\Gamma \otimes \Gamma^\top(L_H \otimes I_{d_Y})^\top\\& \quad \quad \cdot(M\Sigma_y + \Phi\Sigma_xC^\top)F^\top L)\boldsymbol K^{(d_Y, d_Y)}\Bigl) 
\end{align*}
where $\boldsymbol K^{(d_Y, d_Y)} = \sum_{k=1}^{d_Y} \sum_{l=1}^{d_Y} e_{d_Y, l}e_{d_Y,k}^\top \otimes e_{d_Y, k}e_{d_Y,l}^\top$ is the commutation matrix of dimension $d_Y \times d_Y$. Then,
\begin{align*}
    &2\trace \Bigl(\VEC(\tilde E) \VEC(\tilde E)^\top L^\top(\Gamma \otimes \Gamma^\top(L_H \otimes I_{d_Y})^\top\\& \quad \quad \cdot(M\Sigma_y + \Phi\Sigma_xC^\top)F^\top L)\boldsymbol K^{(d_Y, d_Y)}\Bigl) \\
    &= \sum_{i,j=1}^H \sum_{k,l=1}^{d_Y}2\trace \Bigl(\VEC(\tilde E) \VEC(\tilde E)^\top\Bigl((e_i^\top \otimes I_{d_Y})\Gamma(e_j \otimes I_{d_Y})\\ &e_{d_Y,l}e_{d_Y,k}^\top \!\otimes \!(e_i^\top \otimes I_{d_Y})\Gamma^\top(L_H \otimes I_{d_Y})^\top(M\Sigma_y \!+\! \Phi\Sigma_xC^\top)\\
    &F^\top(e_j \otimes I_{d_Y})e_{d_Y,k}e_{d_Y,l}\top\Bigl)\Bigl)  \\
    &= \sum_{i,j=1}^H \Omega\Bigl((e_i^\top \otimes I_{d_Y})\Gamma(e_j \otimes I_{d_Y})J_{d_y}, \\
    & \quad \quad (e_i^\top \otimes I_{d_Y})\Gamma^\top(L_H \otimes I_{d_Y})^\top \\
    &\quad \quad (M\Sigma_y + \Phi\Sigma_xC^\top)F^\top(e_j \otimes I_{d_Y})J_{d_Y}\Bigl)
\end{align*}
where the final equality follows from \Cref{lemma: single-step misspecified}.

Thus, we reach our conclusion with
\begin{align*}
        &\lim_{N\to \infty} N \E \brac{\varepsilon_N } \\
        &= \sum_{i,j=1}^H\Omega\Bigl((e_i^\top \otimes I_{d_Y})F\Sigma_yF^\top(e_j \otimes I_{d_Y}), \\
        & \quad \quad(e_i^\top \otimes I_{d_Y})\Gamma^\top\Gamma(e_j \otimes I_{d_Y})\Bigl) \\
        &\quad \quad +\Omega\Bigl((e_i^\top \otimes I_{d_Y})\Gamma(e_j \otimes I_{d_Y})J_{d_y}, \\
    & \quad \quad (e_i^\top \otimes I_{d_Y})\Gamma^\top (L_H \otimes I_{d_Y})^\top \\
    &\quad \quad (M\Sigma_y + \Phi\Sigma_xC^\top)F^\top(e_j \otimes I_{d_Y})J_{d_Y}\Bigl) \\
    &\triangleq \Theta.
    \end{align*}

\end{proof}

\end{document}